\documentclass[usenatbib]{mn2e}
\usepackage{graphicx}

\title[Neutron-star core-quakes]{Mixed-phase induced core-quakes and the changes in neutron star parameters}
\author[M. Bejger, P. Haensel and  J.L. Zdunik]
{M. Bejger\thanks{e-mail: bejger@camk.edu.pl},
P. Haensel\thanks{e-mail: haensel@camk.edu.pl} and
J. L. Zdunik\thanks{e-mail: jlz@camk.edu.pl}\\
N. Copernicus Astronomical Center, Polish Academy of
Sciences, Bartycka 18, PL-00-716 Warszawa, Poland}
\begin{document}

\date{Accepted: 17 February 2005, Received: 30 December 2004}

\pubyear{2005}
\maketitle

\begin{abstract} We present approximate formulae describing the changes
in neutron star parameters caused by the first-order phase
transition to an ``exotic'' state (pion or kaon condensate, quark matter)
resulting in formation of a mixed-phase core. The
analytical formulae for the changes in radius, moment of inertia
and the amount of energy released during the core-quake are
derived using the theory of linear response of stellar structure
to the core transformation. Numerical coefficients in these
formulae are obtained for two realistic EOSs of dense matter. 
The problem of nucleation of the exotic phase as well as possible
astrophysical scenarios leading to a core-quake phenomenon is also
discussed.
\end{abstract}


%
\section{Introduction}
\label{intro}

After the discovery of neutron stars it became immediately clear
that they are unique  playgrounds to study the properties of
super-dense matter in the most extreme physical conditions. On the
other hand, the very existence of neutron stars, and their
participation in various astrophysical phenomena,  represented a
major challenge for theorists. In order to construct neutron star
models, one needs to know the equation of state (EOS) and other
properties of matter at densities $\ga 10^{15}~{\rm g~cm^{-3}}$.

One of the most intriguing predictions of some theories of dense
matter is a possibility of a phase transition into an ``exotic''
state. Several phase
transitions are predicted by dense-matter theories, including pion and
kaon condensation, and de-confinement of quarks (for review see
e.g. \citealt{weber.book}). The most interesting case, from the point of
view of its consequences for neutron star structure, corresponds
to a phase transition which is of the first-order type. In such a
case, equilibrium phase transition from the normal, lower density
phase to the pure exotic one, occurs at a well defined pressure,
$P_0$,  and is accompanied by a density jump at the phase
interface.

A first order phase transition allows for a metastability of the
pure normal phase with pressure $P>P_0$. Therefore, a metastable
core could form during neutron-star evolution (accretion, spin
down), and nucleation of the exotic phase would then lead to
formation of a new-phase core, accompanied by a core-quake, energy
release, and other phenomena, such as shrinking of the radius and
a speed up of rotation.  Usually, the radius of the new-phase core is
significantly smaller than the original stellar radius, and the
effect of core-quake can be described within the linear response
theory developed by \citet{fop1} and \citet{fop2}; 
an earlier Newtonian theory was presented by \citet{SHZ1983}.
Expressions for the changes of neutron-star
radius, moment of inertia, and energy release were obtained in
these papers in terms of expansions in powers of the  radius of
the new-phase core, the numerical coefficients of the expansions
being determined by the EOS of normal phase, the mass of the
metastable configuration and  the density jump at the phase
transition.

Relaxing  the condition of local (microscopic) electrical
neutrality opens a possibility  of coexistence of two phases of
dense matter, within a range of pressures, in the form of a
mixture of the two phases, lower-density (normal - N) and
higher-density (superdense - S), each of them charged, and the
mixture being electrically neutral on the average only
\citep{Glend91,Glend92}.
The volume fraction occupied by the higher-density
phase grows from zero at the lower pressure boundary, $P^{\rm
(m)}_{\rm N}$,  of the mixed phase up to one
 at the upper pressure boundary, $P^{\rm (m)}_{\rm S}$. If
 the surface tension at the N-S  phase interface is not too large,
 then the mixed phase is preferred over a pure phase state (the latter
 is: N-phase for $P<P_0$ and S-phase for $P>P_0$).

 In this paper paper we study the changes of the neutron-star
 parameters implied by the appearance of an exotic phase which
 forms a mixed-phase core.
 Some astrophysical scenarios which
 could lead to formation of a mixed-phase core were already
 considered in literature (see e.g.
\citealt{GlendPW97,HeiselHjorth1998,Chubarian2000,SpyrSterg2002}).
The phase transition in stellar core
 results from the increase of the central pressure due to
 the pulsar slow-down or mass accretion on the neutron star in a binary system.
 In all these papers the  authors assume that matter remains
 in equilibrium during neutron star evolution, and therefore they use
 an EOS of matter in ground state. However, in a
 first order phase transition, the new phase can appear
 only via nucleation. Therefore, the star acquires first a metastable
 core of normal phase,  in which an exotic phase nucleates. Nucleation
 destabilizes stellar configurations,  implies a core-quake, and
 finally new equilibrium is reached with a core of a
 mixed or a pure exotic phase, the actual outcome depending on the
 detailed kinetics of the first-order phase transition.

In the present paper we derive the formulae for the changes of neutron star
 parameters resulting from the formation of a mixed-phase core. 
These changes are proportional
 to the specific powers of the ratio of the mixed-phase core radius 
to the radius of the last stable pure N-phase configuration.
 We neglect the effects of rotation on neutron-star structure. 

The plan of the article is as follows. In Sect.\ \ref{theory}, our
notation, construction and description of the EOS with a mixed
phase segment is presented. General lowest-order formulae for
the changes in neutron-star parameters are derived in Sect.\ \ref{linear}.
Numerical calculations are performed for two recent realistic
EOSs of the normal phase, in which the exotic dense phase is
assumed to nucleate. Results for the linear response
parameters are presented in Sect.\ \ref{sly} where we also
provide with  some numerical estimates of expected effects.
 In Sect.\ \ref{sect:astro.scen}, we briefly describe the
astrophysical scenarios which could lead to a nucleation of an
exotic phase and a core-quake. Some problems connected with
nucleation  and formation of a mixed-phase core are discussed
in Sect.\ \ref{sect:exotic-nucleation.mixed}. Finally, in
Sect.\ \ref{conclusions} we summarize our results and  we also
discuss their possible applications, and perspectives of their
generalization to rotating neutron stars.
%
\section{EOS with a mixed-phase region}
\label{theory}
%
%
Let us consider a general case of a first-order phase transition
between the N and S phases of dense matter.  As shown by
\citet{Glend91,Glend92}, relaxing the microscopic
charge-neutrality condition can make a mixed-phase state
energetically preferred provided the surface tension and Coulomb
contributions are sufficiently small.

Assume the thermodynamic equilibrium of a multi-component and
multi-phase dense matter, neglecting the Coulomb and surface
contributions to the thermodynamic quantities. The elementary
constituents of the matter are hadrons ($h$), which may be
baryons, quarks, and strongly-interacting meson (pion and kaon)
condensates, as well as  leptons (electrons and muons). The energy
densities in both phases  depend
on the number densities of the matter constituents in these
phases,
\begin{equation}
{\cal E}^{\rm N}={\cal E}^{\rm N}(\lbrace{ n_h^{\rm N}\rbrace},
n_e^{\rm N},n_\mu^{\rm N})~,~~~ {\cal E}^{\rm S}={\cal E}^{\rm
S}(\lbrace{ n_h^{\rm S}\rbrace}, n_e^{\rm S},n_\mu^{\rm S})~.
\label{eq:exotic-E.A.B.phases}
\end{equation}
As the translational invariance may be broken within a phase, the
number densities are actually the volume-averaged ones. We assume
that the size of the region occupied by a non-uniform phase is
sufficiently  large compared to the characteristic length-scale of
the non-uniformity. Then  the volume averages within each phase
are well defined.

The corresponding electric-charge densities (in the units of the
elementary charge) and baryon densities (in the units of nucleon
baryon charge) are given by
\begin{eqnarray}
\rho_{\rm el.}^{\rm N}&=&\sum_{h}n_{h}^{\rm N}q_h -n_e^{\rm N}
-n_\mu^{\rm N}~, ~~ n_{\rm b}^{\rm N}=\sum_{h}n_{h}^{\rm
N}b_{h}~,\cr\cr \rho_{\rm el.}^{\rm S}&=&\sum_{h}n_{h}^{\rm S}q_h
-n_e^{\rm S} -n_\mu^{\rm S}~, ~~n_{\rm b}^{\rm
S}=\sum_{h}n_{h}^{\rm S}b_h~. \label{eq:exotic-rho_e.n_b.A.B}
\end{eqnarray}
where $q_h$ and $b_h$ are, respectively, the  electric and
baryonic charges of a hadron $h$.

 The thermodynamic equilibrium of a mixture of phases N and S, at a fixed average
baryon density $n_{\rm b}$, will be calculated by minimizing the
average energy density
\begin{equation}
{\cal E}=(1-\chi){\cal E}^{\rm N}+ \chi {\cal E}^{\rm S}~.
\label{eq:exotic-E.A.B.chi}
\end{equation}
under the condition
\begin{equation}
n_{\rm b}=(1-\chi)n_{\rm b}^{\rm N}+ \chi n_{\rm b}^{\rm S}~,
\label{eq:exotic-n_b.A.B.chi}
\end{equation}
and under the constraint of  average (macroscopic) electrical
neutrality
\begin{equation}
\rho_{\rm el.}=(1-\chi)\rho_{\rm el.}^{\rm N}+ \chi \rho_{\rm
el.}^{\rm S}=0. \label{eq:exotic-rho_e.A.B.chi}
\end{equation}

For example,  let us consider a first-order phase transition implied by  kaon
condensation. Let us assume that the characteristic length-scale of the hadron
electric-charge inhomogeneities is much smaller than the electron and muon
screening lengths (this is not always true, see e.g. \citealt{NorsenReddy2001}).
Then the electron and muon densities can be considered as uniform,
\begin{equation}
n_e^{\rm N}=n_e^{\rm S}\equiv n_e~,~~~ n_\mu^{\rm N}=n_\mu^{\rm
S}\equiv n_\mu~. \label{eq:exotic-n_mu.n_e.const}
\end{equation}
We have to determine the values of eight variables (four nucleon
densities, two lepton densities, kaon density, and the volume
fraction $\chi$) by minimizing ${\cal E}$, Eq.\
(\ref{eq:exotic-E.A.B.chi}),  under the conditions given by Eqs.\
(\ref{eq:exotic-n_b.A.B.chi}) and (\ref{eq:exotic-rho_e.A.B.chi}).
This leads to a set of nonlinear equations relating the
thermodynamic variables, each of these equations  having a clear
physical meaning. Mechanical equilibrium between the two phases
requires (we remind that surface and  Coulomb contributions are
neglected)
\begin{equation}
P^{\rm N}=P^{\rm S}~. \label{eq:exotic-P.A.B}
\end{equation}
The strong interactions imply the equality of chemical potentials
of nucleons in the two phases,
\begin{equation}
\mu^{\rm N}_n=\mu^{\rm S}_n=\mu_n~,~~~~
\mu^{\rm N}_p=\mu^{\rm S}_p=\mu_p,
\label{eq:exotic-mu.n.p.A.B}
\end{equation}
Finally, weak interactions involving hadrons and leptons lead to
\begin{equation}
\mu_n=\mu_p+\mu_e~,~~~\mu_e=\mu_{K^-}~,~~~\mu_\mu=\mu_e~.
\label{eq:exotic-weak.mixed.Kaon}
\end{equation}
Together with Eqs.\ (\ref{eq:exotic-n_b.A.B.chi}) and
(\ref{eq:exotic-rho_e.A.B.chi}), we get eight equations for eight
thermodynamic variables. The solution corresponds to the
thermodynamic equilibrium at a fixed $n_{\rm b}$ and under the
constraint of macroscopic electrical  neutrality.
\begin{figure}
\centering \resizebox{3.25in}{!} {\includegraphics[clip]{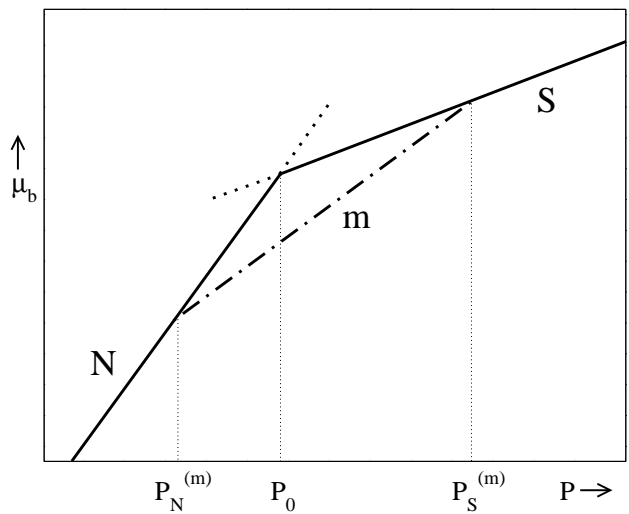}}
\caption{Baryon
chemical potential $\mu_{\rm b}$ versus pressure $P$ in the presence of an
equilibrium  first order phase transition, at $P=P_0$,
 between the N and S phases. The N phase, if  stable with respect
 to the phase transition into the S-phase,  is represented
 by the solid line; its dotted continuation corresponds to an
 {\it over-compressed } state,  metastable with respect to the transformation
 into the S phase.
 Analogous notation is used for the S phase, where the dotted
 continuation of the S-line corresponds to a metastable
 \textit{under-compressed} state.
 The mixed `m'  phase ($P^{\rm (m)}_{\rm N}<P<P^{\rm (m)}_{\rm S}$)
 is represented  by the dash-dotted line.}
\label{fig:mubP.mixed}
\end{figure}
\begin{figure}
\centering \resizebox{3.25in}{!} {\includegraphics[clip]{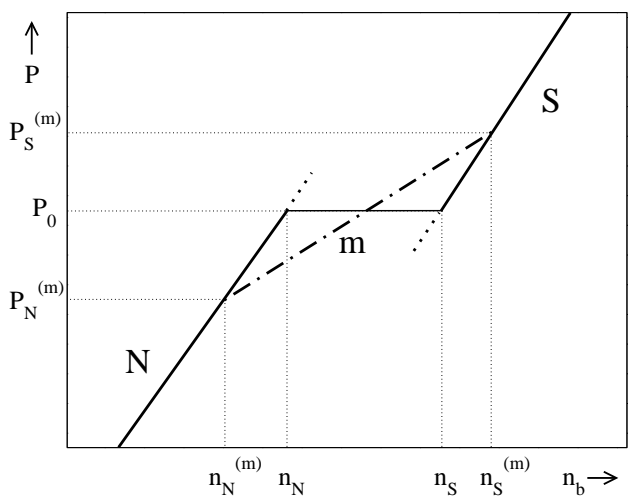}}
\caption{Same as
in Fig.\ \ref{fig:mubP.mixed} but for $P$ versus $n_{\rm b}$. The mixed 
phase ($P^{\rm (m)}_{\rm N}<P<P^{\rm (m)}_{\rm S}$) is represented by the 
dash-dotted line. For further
 explanations see the text.}
\label{fig:Pnb.mixed}
\end{figure}
Let us first discuss the character of the bulk equilibrium as a function of the
pressure, Fig.\ \ref{fig:mubP.mixed}. For $P<P^{\rm (m)}_{\rm N}$,
the equilibrium is realized by the pure N-phase. For $P^{\rm
(m)}_{\rm N}<P<P^{\rm (m)}_{\rm S}$, the equilibrium corresponds
to a mixed m-state of both phases (the m-phase). The volume fraction occupied
by the S-phase increases monotonously with $P$, from zero at
$P=P^{\rm (m)}_{\rm N}$, to one  at $P=P^{\rm (m)}_{\rm S}$. For
$P>P^{\rm (m)}_{\rm S}$ we have a pure S-phase. The calculations
of a mixed-phase state, where  kaon-condensed matter coexisted
with baryon matter, were performed by
\citet{GlendSchaff1998,GlendSchaff1999} and \citet{NorsenReddy2001}.
The models  of dense matter with a mixed phase of de-confined quark
matter coexisting with baryon phase were constructed by
\citet{HeiselPethStaub1993} and \citet{GlendPei95}.
The importance  of
the phase-interface (surface tension and curvature energy) effects
for forming  the mixed-phase state and for its spatial structure  was
studied by
\citet{HeiselPethStaub1993,ChristGlend97,ChristGlendSB2000} and
\citet{NorsenReddy2001}.
A mixed-phase state affects the EOS of dense matter as visualized
in Fig.\ \ref{fig:Pnb.mixed}. For the sake of comparison, we show
also  a standard first-order transition between the pure N and S
phases. For a pure N-S phase transition, the densities $n_{\rm
N}<n_{\rm b}<n_{\rm S}$ could not exist in the stellar interior
because  $P$ should be monotonous there to create an outward
directed force which balances at each point the gravitational
pull.  On the contrary, the mixed-phase layer of density $n^{\rm
(m)}_{\rm N}<n_{\rm b}<n^{\rm (m)}_{\rm S}$ can well exist in the
stellar interior, with the pressure increasing from $P^{\rm
(m)}_{\rm N}$ at the top of the layer to $P^{(m)}_{\rm S}$ at its
bottom. Mixing of the N and S phases softens the EOS as compared
to the pure N-phase EOS, but the effective softening is weaker
than in the limiting case of a pure-phase transition between the N
and S phases.

The surface and Coulomb effects bring positive contributions to
${\cal E}$ and $\mu_{\rm b}$. They affect the  size  and shape of
the structures within the   mixed-phase layer. For a periodic
structure, the virial theorem (see, e.g., \citealt{PethRav1995}
and references therein) tells us that the surface
contribution is twice the Coulomb one. Both are pushing up the
value of $\mu_{\rm b}^{\rm (m)}(P)$. They are particularly
important at the edges of the mixed-phase region, where  the
droplets of  one phase within the dominating one  are small. It is
clear that these effects increase $P_{\rm N}^{\rm (m)}$ and
decrease $P_{\rm S}^{\rm (m)}$, narrowing the mixed-phase layer in
the stellar interior. If sufficiently large, the surface and
Coulomb contributions can entirely remove the mixed phase: the
difference in $\mu_{\rm b}(P)$ of the pure and mixed phases is
usually small. It will occur if surface tension is greater 
than some critical value ($\sigma>\sigma_{\rm crit}$)
so that $\mu_{\rm b}^{\rm
(m)}(P)>\mu_{\rm b}^{\rm N}(P)$ for $P<P_0$ and $\mu_{\rm b}^{\rm
(m)}(P)>\mu_{\rm b}^{\rm S}(P)$ for $P>P_0$. \citet{HeiselPethStaub1993} 
obtained  $\sigma_{\rm crit}\simeq 70~{\rm MeV~fm^{-2}}$
for a transition from nucleon (N) to quark (S) matter.  The actual
value of the surface tension for the quark-matter droplets in
baryonic medium is very poorly known, $\sigma = (10-100)~{\rm
MeV~fm^{-2}}$.

\section{Changes of the stellar parameters for the mixed-phase transition -
linear approach} \label{linear} 
\begin{figure}
\centering
\resizebox{3.25in}{!}{\includegraphics[clip]{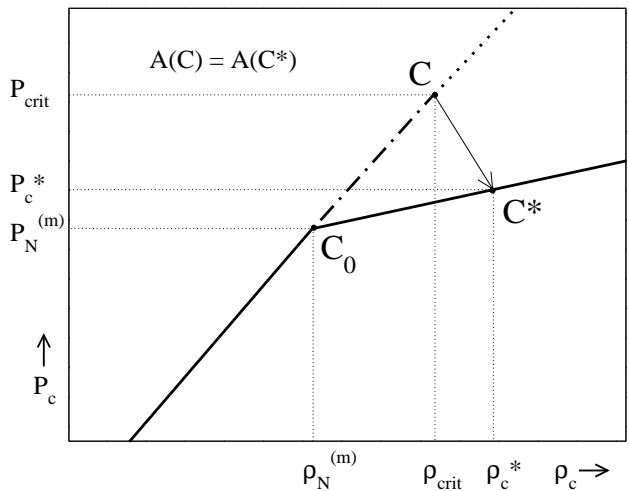}}
\caption{Central pressure  $P_{\rm c}$ versus
 central matter density $\rho_{\rm c}$ for neutron-star
 configurations based on a pure N-phase EOS and an EOS
 with a mixed-phase segment.  Solid line denotes stable states, 
dash-dotted line - the states which a meta-stable with respect to the
  transition to a mixed-phase state. For a critical central density
$\rho_{\rm crit}$ the S-phase nucleates in the super-compressed core
of configuration ${\cal C}$, this results in a transition
${\cal C}\longrightarrow {\cal C}^*$ into a stable configuration
with a mixed-phase core and central density $\rho^*_{\rm c}$.
Both configurations  ${\cal C}$ and ${\cal C}^*$ have the same baryon number
$A$.}
\label{fig:Pcentr.C0CCstar}
\end{figure}

\begin{figure}
\centering
\resizebox{3.4in}{!}{\includegraphics[clip]{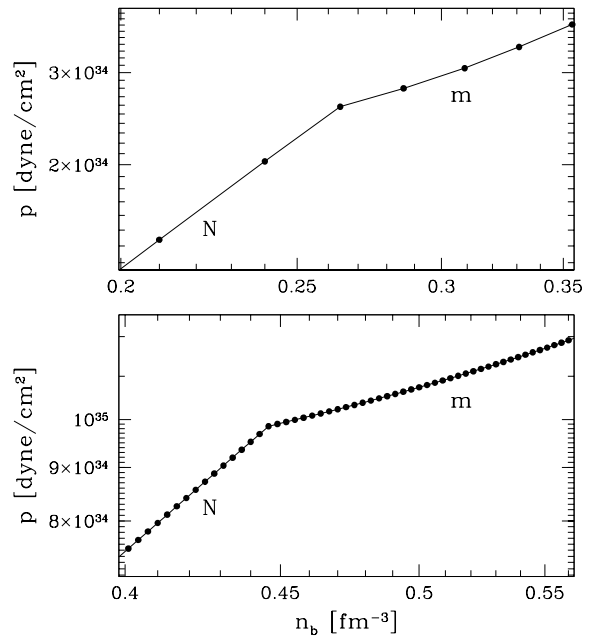}}
\caption{Upper panel: logarithm of pressure $P$ versus the
logarithm of baryon density $n_{\rm b}$ for sample mixed-phase
EOS (quark matter - baryon matter) from \citet{Glend.book},
Table 9.1. The filled circles correspond to the tabulated
points.
The discontinuous drop in  $\gamma$ at the N-m interface is
clearly seen. Notice that the deviation from the linearity of
the EOS of the shown  m-phase  segment in the ${\rm
log}(n_{\rm b})-{\rm log}P$ plane, is small. This means, that
within the considered density interval, which extends from
$0.26~{\rm fm}^{-3}$ to $0.35~{\rm fm}^{-3}$, polytropic
approximation for the m-phase is good. Lower panel: similar
plot for the  GM-GS model with kaon optical potential $U^{\rm
lin}_K=-120~$MeV, producing the EOS with mixed phase segment
(kaon-condensate - baryon matter), obtained by
\citet{Pons2000}. The filled circles correspond to the
tabulated points. The polytropic approximation for the
mixed-phase with kaon condensate appears to be very good.}
\label{exmpleos_lgplgnb}
\end{figure}

We assume that at a central
pressure $P_{\rm c}=P_{\rm crit}$ the nucleation of the
S-phase in a super-compressed core, of radius $r_{\rm N}$, of
configuration ${\cal C}$, initiates  the phase transition and
formation of mixed-phase core of radius $r_{\rm m}$ in a new
configuration ${\cal C}^*$, as presented on Fig.\
\ref{fig:Pcentr.C0CCstar}. Transition to a mixed phase
occurring at $r_{\rm m}$ is associated with a substantial drop
in the adiabatic index of matter, defined as $\gamma\equiv
(n_{\rm b}/P){\rm d}P/{\rm d} n_{\rm b}$, from $\gamma_{\rm
N}$ to $\gamma_{\rm m}$: mixed phase is much softer than the
pure one. This is because in the mixed phase the increase of
mean density is reached partly via  conversion of a less dense
N phase into denser S phase, and therefore requires less
pressure than for a pure phase. The effect exists for any
fraction of the S phase, even in the limit of the fraction of
the S-phase tending to zero, and leads to a discontinuity of
$\gamma$ at $\rho^{\rm (m)}_{\rm N}\equiv \rho_{\rm m}$. An
example of a very dramatic drop of $\gamma$ at $\rho_{\rm m}$
is given in Fig. 20 of \citet{APR98}. Other examples are given
by \citet[][see Fig. 1 in their article]{Pons2000}, 
where the transition from the
solid line (no kaon condensate) to dashed line representing
mixed phase is clearly accompanied by a drop in the derivative
of pressure with respect to density. 
In Fig.~\ref{exmpleos_lgplgnb} we show two examples of realistic
EOSs with a mixed phase segment, one by \citet[][Table 9.1]{Glend.book} 
and the other by \citet[][GM-GS-U120]{Pons2000}. As one can see, 
near the mixed-phase transition point the behaviour of the logarithm of
pressure is linear with respect to the logarithm of baryon
density; it clearly means that the polytropic approximation is
valid in the small mixed-phase core regime.
This is precisely the case we are interested in 
- we will henceforth benefit from expansion in powers of the 
core radius $r_{\rm m}$.
The validity of the $\gamma_{\rm m}=const.$ approximation in
the limit of small $r_{\rm m}$ can be also phrased 
differently, using the small-$r_{\rm m}$ expansions.
Using
techniques described in detail in \citet{fop2}, we can see
that the difference between the value of $\gamma$ at
$\rho_{\rm m}$, and at the center of the ${\cal C}^*$
configuration in Fig.~\ref{fig:Pcentr.C0CCstar} is quadratic in $r_{\rm
m}$, so that $\gamma^*_{\rm c}=\gamma^{\rm (m)}(\rho_{\rm
c}^*)=\gamma_{\rm m}+{\cal O}(r_{\rm m}^2)$. As we show below,
the leading changes due to a core-quake are proportional to
the fifth (radius, moment of inertia) and seventh
(gravitational mass) power of $r_{\rm m}$. They are obtained
by putting $\gamma^*_{\rm c}\simeq \gamma_{\rm m}$. Inclusion
of quadratic correction in the expression for $\gamma^*_{\rm
c}$ contributes to the higher order terms, i.e., seventh power
of $r_{\rm m}$ in the case of radius and moment of inertia,
and ninth power of $r_{\rm m}$ in the case of gravitational
mass.

Having a pair of EOSs, one with and one without a
mixed-phase softening, we compare the hydrostatic equilibria
of neutron stars corresponding to each of these  EOSs. The
models which we calculate  are non-rotating, spherically symmetric
solutions of Einstein's equations, which in this special case reduce to the
Tolman-Oppenheimer-Volkoff equations \citep{Tolman39,TOV}.
The neutron-star models are labeled by the
central density $\rho_{\rm c}$.
The configurations calculated with these two
EOSs are identical up to $\rho_{\rm c}=\rho^{\rm (m)}_{\rm N}$; the
configuration with such central density is denoted by ${\cal C}_0$, and
will be treated as a ``reference configuration''.

 When the central density exceeds
$\rho^{\rm (m)}_{\rm N}$,  the models based on these EOSs are
different,  due to the appearance of a softer mixed-phase core in configurations
corresponding to the mixed-phase segment of the EOS.
For a fixed baryon number $A$, greater than a
baryon number $A_0$ of  the reference configuration ${\cal C}_0$,
 we compare the global parameters, such as  mass-energy, radius,
 and moment of inertia. Their difference corresponds to the changes in these
 parameters implied by the phase transition in the stellar core.

From now on we will restrict ourselves to a linear response of
neutron star to the appearance of the mixed phase core. The
calculation  is based on expressing the change in the density
profile, due to the presence to a small core, as the combination
of two independent solutions of the linearly perturbed equations
of stellar structure \citep{fop1,fop2}.
The presence of a denser phase in the core changes the boundary
condition at the phase transition pressure $P^{\rm (m)}_{\rm N}$
and allows us to determine the numerical coefficients in the
expression for the density profile change. The leading term in the
perturbation of the boundary condition at the edge of the new
phase results from the mass excess due to the lower stiffness, and
higher density of the new phase as compared to those of the
N-phase.

Let us compare basic expressions obtained when the new core
consists of a pure S-phase, considered by \citet{fop1}
and \citet{fop2}, and in the present case of a mixed-phase
core. For a  pure S-phase core the phase transition is accompanied 
by density jump $\rho_{\rm N} \to
 \rho_{\rm S}$ which 
leads to the lowest-order expression for the
core-mass excess (with respect to the pure N-phase configuration),
\begin{equation}  \label{deltamfo}
\delta m_{\rm core} = {4\over3}\pi (\rho_{\rm S}-\rho_{\rm N}) r_{\rm S}^3
 + \mathcal{O}(r_{\rm S}^{5}),
\end{equation}
where powers $(r_{\rm S})^l$ with $l>3$ have been neglected. In
the case of a  mixed-phase core, considered in the present paper,
the lowest-order expression vanishes,  because there is no density
jump at the core radius.

Let us introduce the notation appropriate for description of the change
of the properties of matter at the N-phase -- m-phase transition point.
The transition occurs at $P=P_{\rm m}\equiv P^{\rm (m)}_{\rm N}$. Density
does not change and is equal $\rho_{\rm m}\equiv
\rho^{\rm (m)}_{\rm N}$, while the adiabatic
index changes from $\gamma_{\rm N}$ on the N-side to a lower value
 $\gamma_{\rm m}$ on the m-side.

To the lowest order in $r_{\rm m}$,  the mass excess in our case
 is  due to the difference in the stiffness (adiabatic index) of the matter
in N-phase and m-phase. Expanding term by term, we get
\begin{equation}  \label{deltamso}
\delta m_{\rm core} = {4\pi\rho_{\rm m}\over45}(1+x_{\rm m})(1+3x_{\rm m})
(\kappa_{\rm m}^2-\kappa_{\rm N}^2)
r_{\rm m}^{5} + \mathcal{O}(r_{\rm m}^{7})
 \end{equation}
where $x_{\rm m} $ is the ratio of pressure $P_{\rm m}$, and the
energy density $\rho_{\rm m} c^2$ at the phase transition point,
$x_{\rm m}=P_{\rm m}/\rho_{\rm m}c^2$.
The parameters $\kappa_{\rm N}^2$, $\kappa_{\rm m}^2$ are defined
by:
\begin{equation}  \label{kappa}
\kappa_{\rm N}^2=4\pi G \rho_{\rm m}/v_{\rm N}^2, ~~\kappa_{\rm
m}^2=4\pi G \rho_{\rm m}/v_{\rm m}^2~,
\end{equation}
where $v_{\rm N}$ and $v_{\rm m}$ denote the sound velocity on
the both sides of the core boundary. Let us remind that
sound velocity $v$ is related to the adiabatic index by
$v^2= \gamma P/(\rho+P/c^2)$.

The main parameter of the linear theory - the radius of the core of mixed phase
$r_{\rm m}$ is connected with the density range
 of metastability which can be achieved by the
N phase of matter, i.e., the difference between $\rho_{\rm crit}$ and
$\rho_{\rm N}^{\rm (m)}$, via relation (see \citealt{fop2}):
\begin{equation}  \label{rhorm}
\Delta \bar\rho_{\rm crit} := {\rho_{\rm crit}-\rho_{\rm N}^{\rm (m)}\over \rho_{\rm N}^{\rm (m)}}=
{1\over6}\kappa_{\rm N}^2(1+x_{\rm m})(1+3x_{\rm m})r_{\rm m}^{2}.
\end{equation}

Strictly speaking, the lowest order works only for sufficiently
small value of $r_{\rm m}$. From Eq.\ (\ref{deltamso}) and 
Eq.\ (\ref{kappa}) we have 
\begin{eqnarray}\label{deltamso1} 
\delta m_{\rm core} = {4\over45}\pi \rho_{\rm m}(1+x_{\rm m})^2 (1+3x_{\rm m})
  ~~\times \nonumber &
  \\
  \times~~{4\pi
G\rho_{\rm m}^2\over \gamma_{\rm N}P_{\rm m}}
(\gamma_{\rm N}/\gamma_{\rm m}-1)\cdot
{r_{\rm m}}^5 + \mathcal{O}(r_{\rm m}^7)
  \end{eqnarray}

The new phase manifests itself in the boundary conditions by the
presence of the prefactor $(\gamma_{\rm N}/\gamma_{\rm m}-1)$,
which acts similarly as the prefactor $(\rho_{\rm S}/\rho_{\rm
N}-1)$ in the case of a first order phase transition from pure
N-phase to pure S-phase. The fact that in our case the density
stays continuous, results in the linear-response effects which
are proportional to a power of the core radius greater by two than in
the case of the transition to a pure S-phase.

On the basis of the above analytical
considerations, we expect that the relative changes in stellar parameter
 $Q$ contain a prefactor $(\gamma_{\rm N}/\gamma_{\rm m}-1)$  and
 are proportional to $r_{\rm m}^5$ in the case of $Q=R,I$ and to
 $r_{\rm m}^7$ in the case of mass-energy $E=Mc^2$. Summarizing, we
 expect the following form of the lowest-order expressions:
\begin{equation}  \label{beta1}
\delta\bar{Q} \equiv \frac{Q^*-Q}{Q_0} \simeq
-(\gamma_{\rm N}/\gamma_{\rm m}-1)\cdot\beta_{Q}
\cdot ({\bar{r}_{\rm m}})^l,
\end{equation}
where $\bar{r}_{\rm m}\equiv r_{\rm m}/R_0$, and  $l = 5$ for
radius $R$ and moment of inertia $I$, and $l = 7$ for the energy
$E$. The coefficients $\beta_Q$ are functionals of the reference
configuration, $\beta_Q=\beta_Q[{\cal C}_0]$.
%
\section{Changes in stellar parameters for realistic EOSs}
\label{sly}
%
In order to explore how a realistic neutron star will react on
the appearance of a mixed-phase core, to describe the N-phase
we used two recent EOSs: SLy of \citet{sly} and FPS EOS of
\citet{FPS}. Both EOSs  describe in unified way (i.e.,
starting from a single effective nuclear Hamiltonian) both the
crust and the core of neutron star. They assume that neutron
star core is composed of neutrons, protons, electrons and
muons.

In the actual calculations, we approximated the mixed-phase
softening by replacing the mixed-phase segment of the EOS
starting at $n_{\rm m}\equiv n^{\rm (m)}_{\rm N}$, by a
polytrope with $\gamma_{\rm m}<\gamma_{\rm N}$.  As we
explained in Sect. 3, this  is equivalent to neglecting the
r-dependence of $\gamma$ within the mixed-phase core, and it
does not change the leading terms of the small $r_{\rm
m}$-expansions. Also, as shown in Figs.~\ref{exmpleos_lgplgnb}, 
this approximation is excellent for not too large
mixed-phase core.

We constructed a large set of EOSs with a mixed-phase segment,
using many values of the phase-transition density $n_{\rm m}$
and  several values of $\gamma_{\rm m}$. In this way we were
able to study the linear response of neutron star to the
appearance of a small mixed-phase core for a wide choice of
the mixed-phase parameters.

Note again that the coefficients $\beta_Q$ depend only on
the properties of the reference configuration ${\cal C}_0$ - 
the EOS of the mixed-phase (here approximated by a polytrope) 
enters merely via $\gamma_{\rm m}$ in the factor
$(\gamma_{\rm N}/\gamma_{\rm m}-1)$ in Eq.\ (\ref{beta1}).

We apply the arguments presented in Sect. \ref{linear} to the
 EOS constructed in the manner described above. By changing
the value of $\rho_{\rm m}$, we obtain  a family of reference
configurations ${\cal C}_0$, and by comparing ${\cal C}$ and
${\cal C}^*$ we extract the values of the linear response
coefficients $\beta_Q[{\cal C}_0]$.
\subsection{Results for SLy EOS of the N-phase}
\label{sect:beta.SLy}
\begin{figure}
\centering
\resizebox{3.25in}{!}{\includegraphics{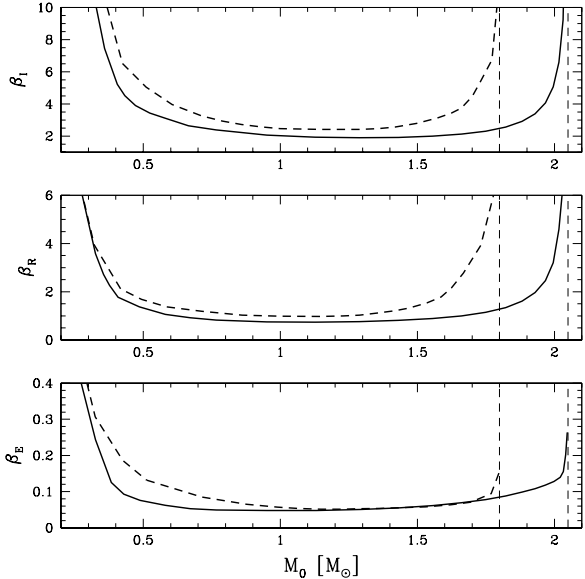}}
\caption{ The linear response parameters $\beta_Q$, versus the
mass $M_0$ of the reference configuration ${\cal C}_0$. The
N-phase is described by the SLy EOS (solid lines, $M_{\rm
max}=2.05~{\rm M}_{\odot}$) and, for comparison, by FPS EOS
(dashed lines, $M_{\rm max}=1.79~{\rm M}_{\odot}$). }

\label{fc0}
\end{figure}

\begin{figure}
\centering \resizebox{3.25in}{!}{\includegraphics[clip]{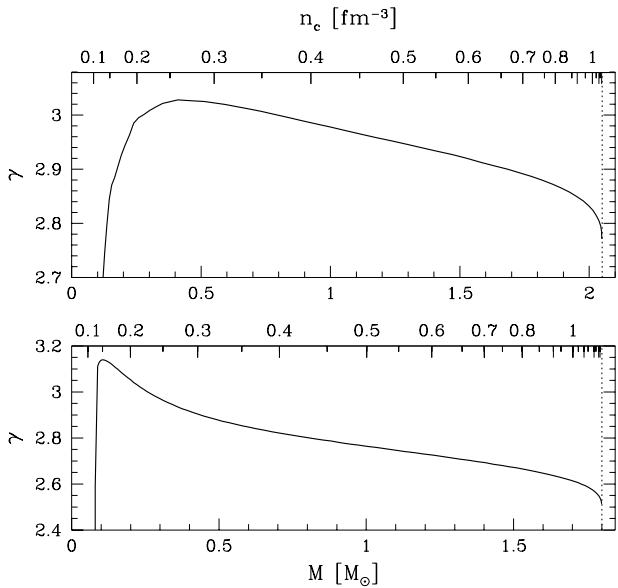}}
\caption{The value of the adiabatic index at the center of non-rotating
configuration for the SLy EOS (upper panel) and FPS EOS (lower panel),  
plotted against the gravitational mass (bottom
horizontal axis), and baryon number density $n_{\rm c}$ at the star center
(upper horizontal axis). Vertical dotted line marks the maximum allowable mass
($M_{\rm max}=2.05~{\rm M}_{\odot}$ for the SLy EOS, and 
$M_{\rm max}=1.79~{\rm M}_{\odot}$ for the FPS EOS). }
\label{slygamma}
\end{figure}

The values of $\beta_Q[{\cal C}_0]$ are  plotted in Fig.
\ref{fc0}. An important parameter relevant for the linear
response to a perturbation of equilibrium configuration is the
adiabatic index of the N-phase  EOS. We plot its value at the
center of the reference configuration ${\cal C}_0$ versus
$M[{\cal C}_0]$ and its central baryon number density in Fig.
\ref{slygamma}.
As it can be seen on Fig. \ref{fc0}, for a wide astrophysically interesting
range of neutron-star masses, the values of the functionals $\beta_Q[{\cal
C}_0]$ stay almost constant.
The plateau values for SLy EOS are $\beta^{\rm plat}_{I}=2.0$,
$\beta^{\rm plat}_{R}=0.8$, and $\beta^{\rm plat}_E=0.05$ for the moment
of inertia, radius and mass-energy, respectively ($\beta^{\rm plat}_{I}=2.4$,
$\beta^{\rm plat}_{R}=1.0$, and $\beta^{\rm plat}_E=0.05$ for FPS EOS, see next
sub-section for details)

The values of $\beta_Q$ increase rapidly with decreasing mass
below $0.4~{\rm M}_\odot$. This is due to the specific generic
features of the realistic EOS of neutron star crust
\citep{PH.Trento2001}, which results in the existence of the
minimum neutron star mass $M_{\rm min}$ (see
\citealt{HZD2002}, and references therein). For
$M\longrightarrow M_{\rm min}$ neutron stars become more and
more ``soft'' with respect to the fundamental mode of radial
perturbation, and become unstable at $M_{\rm min}$. Therefore,
at the same mass-excess $\delta m_{\rm core}$ the effect of
neutron star ``shrinking'' due to an increased gravitational
pull by the mixed-phase core grows, and very rapidly
non-linear effects become important.

The ``numerical plateau'' values $\beta^{\rm plat}_Q$ deserve an
additional comment. The plateau region extends within $0.5{\rm
M}_\odot \la M[{\cal C}_0] \la 0.9M_{\rm max}$.  It is easy to
verify that our plateau values of $\beta_Q$ multiplied by $\simeq
3$ are very close to the linear response coefficients for the pure
S-phase core, $\alpha_Q$, obtained for a medium-stiff EOS of the
N-phase by \citet{fop1}. Clearly, apart from some
normalization constant, the linear response coefficient does not
depend on the nature of the new-phase core, which enters only via
$(\rho_{\rm S}/\rho_{\rm N}-1)(\bar{r}_{\rm S})^l$ in the case of
a first-order phase transition to a pure S-phase, and via
$(\gamma_{\rm N}/\gamma_{\rm m}-1)(\bar{r}_{\rm m})^l$ for a
mixed-phase core. This results from the fact that the change of
the stellar parameters $Q$ depends on the mass excess in the core
via the relation (leading term): $\delta Q \simeq ({\partial Q
/\partial m}) \delta m_{\rm core}$ and the numerical factors
multiplying $(\gamma_{\rm N}/\gamma_{\rm m}-1) (\bar{r}_{\rm
m})^5$ or $(\rho_{\rm S}/\rho_{\rm N}-1)(\bar{r}_{\rm S})^3$ in
$\delta m_{\rm core}$ in the case of second and first order
transition respectively differ by ${1\over 15}\kappa^2_{\rm N}
R^2$ which is approximately 1/3.
\subsection{Results for FPS EOS of the N-phase}
\label{sect:beta.FPS}
The values of $\beta_Q[{\cal C}_0]$ are  plotted in Fig.\
\ref{fc0}, where they can be easily compared with those
obtained for the SLy EOS.  The plateau values are $\beta^{\rm
plat}_{I}=2.4$, $\beta^{\rm plat}_{R}=1.0$, and $\beta^{\rm
plat}_E=0.05$. As we see, they depend rather weakly on the EOS
of the N-phase, provided it is realistic. The $M_0$ range of
the plateau region, $(0.6-1.6)\;M_\odot$, is significantly
narrower than for the SLy EOS. One can try to point out
reasons for this. First, on the high-mass side, the
difference results from the fact that $M^{\rm FPS}_{\rm
max}=1.79~M_\odot$, to be compared with $M^{\rm SLy}_{\rm
max}=2.05~M_\odot$. Otherwise, the quasi-constancy is valid,
similarly as for the SLy EOS, up to about $0.9~M_{\rm max}$,
which seems to be generic.  ``Softness'' of hydrostatic
equilibrium at $M$ close to $M_{\rm max}$ with respect to
radial perturbations is the general-relativistic feature, and
does not depend on the EOS of the N-phase.  On the contrary,
on the low-mass side, the difference in the $\beta_Q$ behavior
reflects the differences between the EOSs at sub-nuclear
densities. This difference between the SLy and FPS EOSs was
studied, albeit in a different context, by \citet{HZD2002}. 
In the relevant stellar mass range, the FPS stars are less bound and
therefore, with decreasing mass, its $\beta$ coefficients
start to increase earlier than for the SLy EOS.
\subsection{Estimating changes in stellar parameters in a
core-quake} \label{changes.num}
Let us now calculate the changes in stellar parameters. First,
the estimates will be obtained for the  SLy EOS of the
N-phase.
Consider ${\cal C}_0$ configuration of mass $1.4~{\rm
M}_\odot$ calculated for the SLy EOS. It has $R\equiv
R_0=11.73$~km, $I\equiv I_0=1.37\times 10^{45}~{\rm g~cm^2}$,
and a central adiabatic index $\gamma_{\rm N}\simeq 2.94$.
Using the plateau values of the $\beta_Q$ coefficients, we can
rewrite the formulae for the changes in neutron-star
parameters in a form suitable for making numerical estimates.
For the stellar radius we get,
%
\begin{equation}
\Delta R\simeq -4.23\times 10^{-5}\cdot(2.94/\gamma_{\rm m}-1)
 (r_{\rm m}/1~{\rm km})^5~{\rm km}.
\label{eq:numer.R}
\end{equation}
%
Assume $\gamma_{\rm m}=1.5$. Then,
 a 1 km core implies shrinking by about 4 cm, a very
small star-quake indeed, but still larger by more than an order of magnitude
than  that associated with macro-glitches in the pulsar timing.
However, the rise of the
star-quake amplitude with $r_{\rm m}$
 is very steep and for $r_{\rm m}=4$~km
we get $\Delta R=42$~m, rather impressive even by the
terrestrial standards.

Expression for the fractional change in $I$ is
%
\begin{equation}
\Delta I/I_0 \simeq -9\times 10^{-6}\cdot(2.94/\gamma_{\rm m}-1) (r_{\rm
m}/1~{\rm km})^5~.
\label{eq:numer.I}
\end{equation}
%
Assume as before that $\gamma_{\rm m}=1.5$.
Formation of a 1 km core implies speed-up of the neutron star rotation
by  $\Delta\Omega/\Omega\simeq -\Delta I/I\simeq 10^{-5}$, one order of magnitude
larger than in the biggest pulsar macro-glitches. If the core radius is
4 km, then $\Delta\Omega/\Omega\simeq -\Delta I/I\simeq 9\times
10^{-3}$, a very distinct feature of a neutron-star core-quake.

Let consider now the energy release. We have
%
\begin{equation}
\Delta E \simeq 4.1\times 10^{45}\cdot(2.94/\gamma_{\rm m}-1)
 (r_{\rm m}/1~{\rm km})^7~{\rm erg},
\label{eq:numer.M}
\end{equation}
%
which means that formation of a 1 km core with $\gamma_{\rm
m}=1.5$ will release a total energy of $4\times 10^{45}$ erg.
Corresponding amount of released energy for a core with $r_{\rm
m}=2~{\rm km}$ and $r_{\rm m}=4~{\rm km}$ is $5\times 10^{47}$ erg
and $6.4\times 10^{49}$ erg, respectively. It may be worthwhile to
compare it with energies associated with a few known astrophysical
processes involving neutron stars. For example, the energy
associated with Soft Gamma Repeaters (SGR) outbursts, which are
now believed to be caused by a extremely strong magnetic field, is
$\sim 10^{44} - 10^{45}$ erg. The Vela pulsar macro-glitches
release from $10^{38}$ to $10^{42}$ erg.  The thermonuclear Type I
X-ray bursts involve  energies   $10^{36}-10^{39}$ erg, but in the
case of the so-called super-bursts the released energy can be as
high as $10^{43}$. By adjusting the core radius, the
 energies associated with core-quakes can be made similar to those released
in phenomena listed above. It should be however noted that the
core radius is not a free parameter in our approach but is
determined by the range of metastability possible in the
considered physical situation which is rather poorly known and
model dependent. The leading terms  in Eqs.\
(\ref{eq:numer.R}-\ref{eq:numer.M})  for the radius $r_{\rm
m}=1~{\rm km}$  correspond to the degree of metastability,
which is usually called ``over-compression''  of the order of
0.5\%, i.e $\Delta \bar\rho_{\rm crit} \simeq 0.005$, see Eq.\
(\ref{rhorm}).

Up to now, we performed all numerical estimates for the SLy
EOS of the N-phase. For the same values of $r_{\rm m}$ and
$\gamma_{\rm m}$, the values of $\Delta R$ etc., for the FPS
model of the N-phase, will be somewhat different, but these
differences are of no practical importance.
What is really crucial, is the strong dependence of $\Delta
R$, $\Delta I$, and especially of $\Delta E$, on the
mixed-phase core radius.
\section{Astrophysical scenarios}
\label{sect:astro.scen}
One can consider several astrophysical scenarios of the formation
of a mixed-phase core in neutron star. In all cases we are dealing
with a two step process. The first step consists in the nucleation
of the S-phase in the metastable N-phase medium, studied by
\citet{HaenSchaeff82,MutoTats1992,IidaSato1997,IidaSato1998}.
The second step consists in the relaxation of
the system to a stable mixed state in which N-phase coexists with
a S-phase.

The first scenario is connected with neutron star birth, in which
proton-neutron star is formed in gravitational collapse of a
massive stellar core. Central core of a proto-neutron star has a
relatively low density, compared to central density of  neutron
star in which it will transform eventually (after a few minutes).
This is due  to large lepton fraction (30-40\%) resulting from
neutrino trapping. The diffusion of neutrinos and resulting
deleptonization of matter is connected with softening of the
neutron-star core EOS and matter compression to higher density.
Simultaneously, the core is heated (up to $30-60~{\rm MeV}/k_{\rm
B}$) by the deposition of the energy of strongly degenerate
electron neutrinos via their down-scattering on the constituents
of dense medium. The  optimal conditions for nucleation of the
S-phase are just after deleptonization (i.e., $\sim$ 10-20 seconds
after the bounce ending the collapse phase). The thermal
fluctuations at $T\sim 40~{\rm MeV}/k_{\rm B}$ are expected to be
sufficient to overcome energy barriers, resulting from the Coulomb
and surface effects,  and to form a mixed-phase core in
thermodynamic equilibrium.

Central compression of matter during slow-down of pulsar rotation is a second
scenario to be considered. Let the pulsar rotation frequency be $\nu$. The
effect of rotation on the pressure distribution is to lowest order $\propto
\nu^2$. Therefore, the rate of the quasi-static increase of central pressure is
$\dot{P}_{\rm c}\propto -\dot{\nu}\nu$. The central-core temperature is
$<10^{9}~$K, so that nucleation can proceed only via quantum fluctuation in
super-compressed core. Nucleation induces pressure deficit, collapse of the
central core accompanied by matter heating, and energy release. During these
violent processes involving matter flow, N and S-phase mix forming a
mixed-phase core.

A third scenario involves neutron star accreting matter in a close
binary system. Central pressure  increases due to the gravity of
accumulated layers of accreted matter, at a rate $\dot{P}_{\rm c}
\propto  \dot{M}$. The core temperature remains below $10^{9}~$K (see
e.g. \citealt{MiraldaHP1990}), and therefore quantum
fluctuations are crucial for the nucleation process to start. Then
the pressure deficit triggers core collapse, accompanied by matter
heating and flow, and a mixed-phase core forms.

\section{Some problems connected with nucleation of a mixed  normal-exotic phase}
\label{sect:exotic-nucleation.mixed}
Let us assume  that the phase transition from a pure N-phase to a
pure S-phase is the first-order one, at $P=P_0$ (Fig.\
 \ref{fig:mubP.mixed}). Let us further assume, that within the
pressure interval $P^{\rm (m)}_{\rm N}<P<P^{\rm (m)}_{\rm S}$ the
thermodynamic equilibrium of dense matter is realized in the form of
the mixed NS phase.

Consider  a neutron star, where the  central pressure $P_{\rm c}$
increases in time  due to a mass accretion in a binary system  or
due to a spin-down of neutron-star rotation. The central pressure
increases on a characteristic timescale  $t_{\rm evol}=P_{\rm
c}/{\dot{P}_{\rm c}}$. Let us assume that initially the stellar
core consisted of a pure N phase. In this case the internal
temperature does not exceed $10^9~$K, the thermal fluctuations are
negligible and the S phase can nucleate only via quantum
fluctuations on a certain timescale $\tau$. 
However, as long as $P_{\rm c}<P_0$, the nucleation
of the S-phase in the quantum regime is impossible
($\tau=\infty$). The actual nucleation will start at some $P_{\rm
c}=P_{\rm nucl}>P_0$, by a formation of a single droplet. If the
expansion rate of the first drop is larger than the  formation
rate of other  droplets in the core, then the pure S phase will
fill the central core with $P\le P_0$. The S-phase core  will then
coexist with the outer layer of the N phase, with a baryon density
drop $n_{\rm S}-n_{\rm N}$ at the interface. Such a scenario seems
likely in the case of the quark-matter nucleation
\citep{IidaSato1997,IidaSato1998}

The central temperature of a proto-neutron star stays high, $T_{\rm
c}>10^{11}~$K,  for some ten seconds. This may be enough to
complete the nucleation of the S phase below $P_0$ and mix it with
the N phase, achieving the thermodynamic equilibrium. It is not
excluded, however, that the core will have no time to evolve to
thermodynamic equilibrium in the low-temperature limit, because of
a too rapid cooling to $T\approx 10^{10}$~K. It is therefore
possible that the final mixed state with  some spatial structure
will be quite different from the strict ground state of the core
at $T=0$; the mixed phase  can remain ``frozen'' in some
metastable state in the core or its part.

Finally, let us mention two difficulties in forming a mixed phase
in  kaon-condensed neutron-star cores \citep{Norsen2002}. It is difficult to
nucleate kaon condensate because of the slowness of weak
interaction processes; very high temperature $T>10^{11}~$ K and low kaon
effective masses are required for the condensation due to thermal
fluctuations. This may happen only in massive newborn neutron
stars, with particularly high central temperatures and densities,
where  a mixed-phase core could be formed. Medium-mass neutron
stars have insufficient central density to nucleate the kaon
condensate at their  birth. On the other hand, high-mass neutron
stars,  which gain their  mass by accretion, can remain in a
metastable state forever, because their internal temperature is
too low for nucleating  kaon condensate in their cores.

%
\section{Conclusions}
\label{conclusions}
%
Formation of a mixed-phase core implies shrinking of neutron star radius,
spin-up, and is associated with energy release. Changes in neutron star
parameters strongly depend on the size of the mixed-phase core. The decrease of
stellar radius, $\Delta R$,  and of the moment of inertia, $\Delta I$,  are
proportional to the fifth power of the core radius. The total energy released
during the core-quake, $\Delta E$,
 is proportional to the seventh power
of the core radius. These powers are higher by two than the
powers of the core radius in the formulae expressing the changes
of the stellar parameters implied by the formation of a pure
high-density phase core.

Apart of the core radius, the formulae involve also the drop
of the adiabatic index at the mixed-phase core edge, and
a numerical coefficient determined by the normal-phase
configuration. In the stellar mass range
$0.5M_\odot\la M \la 0.9M_{\rm max}$ the numerical coefficients
for $\Delta R$, $\Delta I$ and $\Delta E$  depend rather weakly
on $M$ and can be replaced by their  ``plateau value''.

This paper completes the study, within the linear response theory,
 of the effect of phase transitions in the neutron star core on the
stellar structure. It extends the methods  initiated  in the 1980s
for a single (pure) dense phase core \citep{SHZ1983,fop1,fop2}
to the case of a mixed-phase core.
Our results are presented in the form of formulae which are convenient
to make numerical estimates, and which become very precise for small
mixed-phase cores, when the brute-force calculations are impossible
because of limited numerical precision.

The final state reached in a phase transition in neutron-star core
depends on the kinetics of this process. Even if the mixed state is
the stable final configuration, it does not mean that it is reached,
because a one-phase core may form instead. Mixed-state is particularly
difficult to be formed if a significant  strangeness production is
necessary in the dense-phase nucleation process. At temperatures
and densities characteristic of the cores of accreting neutron stars,
the mixed-phase core may never form.
\section*{acknowledgments}
We are grateful to J. Pons for sending
us the tables of his EOSs with the kaon-condensed mixed phase.
This work was partially supported by the KBN grant no.
1P03D-008-27 and by the Astrophysics Poland-France (Astro-PF)
program.


\begin{thebibliography}{}
\bibitem[Akmal, Pandharipande \& Ravenhhall(1998)]{APR98}
Akmal A., Pandharipande V.R., Ravenhall D.G., 1998, Phys. Rev. C, 58, 1804
\bibitem[Christiansen \& Glendenning(1997)]{ChristGlend97}
Christiansen M. B., Glendenning N. K., 1997, Phys. Rev. C, 56, 2858
\bibitem[Christiansen, Glendenning \& Schaffner-Bielich(2000)]{ChristGlendSB2000}
Christiansen M. B., Glendenning N. K., Schaffner-Bielich J., 2000, Phys. Rev. C, 62, 025804
\bibitem[Chubarian et al.(2000)]{Chubarian2000}
Chubarian E., Grigorian H., Poghosyan G., Blaschke D., 2000, A\&A,
357, 968
\bibitem[Douchin \& Haensel(2001)]{sly}
Douchin F., Haensel P., 2001, A\&A, 380, 151
\bibitem[Glendenning(1991)]{Glend91}
Glendenning N. K., 1991, Nucl. Phys. B (Proc. Suppl.), 24B, 110
\bibitem[Glendenning(1992)]{Glend92}
Glendenning N. K., 1992, Phys. Rev. D, 46, 1274
\bibitem[Glendenning(1997)]{Glend.book}
Glendenning N. K., 1997, Compact Stars. Nuclear Physics, Particle
Physics, and General Relativity, Springer, Berlin
\bibitem[Glendenning \& Pei(1995)]{GlendPei95}
Glendenning N. K., Pei S., 1995, in Eugene Wigner Memorial Issue
of Heavy Ion Physics (Budapest), p.1
\bibitem[Glendenning, Pei \& Weber(1997)]{GlendPW97}
Glendenning N. K., Pei S., Weber F., 1997, Phys. Rev. Lett., 79,
1603
\bibitem[Glendenning \& Schaffner-Bielich(1998)]{GlendSchaff1998}
Glendenning N. K., Schaffner-Bielich J., 1998,  Phys. Rev. Lett.,
81, 4564
\bibitem[Glendenning \& Schaffner-Bielich(1999)]{GlendSchaff1999}
Glendenning N. K., Schaffner-Bielich J., 1999,  Phys. Rev. C, 60,
025803
\bibitem[Haensel \& Schaeffer(1982)]{HaenSchaeff82}
Haensel P., Schaeffer R., 1982, Nucl. Phys. A 381, 519
\bibitem[Haensel(2001)]{PH.Trento2001}
Haensel P., 2001, in: Physics of Neutron Star Interiors, eds. D.
Blaschke, N.K. Glendenning, A. Sedrakian, Springer, Berlin
\bibitem[Haensel, Zdunik \& Schaeffer(1986)]{fop1}
Haensel P., Zdunik J. L., Schaeffer R., 1986, A\&A, 160, 251
\bibitem[Haensel, Zdunik \& Douchin(2002)]{HZD2002}
Haensel P., Zdunik J. L., Douchin F., 2002, A\&A, 385, 301
\bibitem[Heiselberg, Pethick \& Staubo(1993)]{HeiselPethStaub1993}
Heiselberg H., Pethick C. J., Staubo  E. F., 1993, Phys. Rev. Lett.,
70, 1355
\bibitem[Heiselberg \& Hjorth-Jensen(1998)]{HeiselHjorth1998}
Heiselberg H., Hjorth-Jensen M., 1998, Phys. Rev. Lett., 80, 5485
\bibitem[Iida \& Sato(1997)]{IidaSato1997}
Iida K., Sato K., 1997,  Prog. Theor. Phys., 98, 277
\bibitem[Iida \& Sato(1998)]{IidaSato1998}
Iida K., Sato K., 1998, Phys. Rev. C., 58, 2538
\bibitem[Muto \& Tatsumi(1992)]{MutoTats1992}
Muto T., Tatsumi T., 1990, Prog. Theor. Phys. 83, 499
\bibitem[Miralda-Escud{\'e}, Haensel \& Paczy{\'n}ski(1990)]{MiraldaHP1990}
Miralda-Escud{\'e} J., Haensel P., Paczy{\'n}ski B., 1990, ApJ 362, 572
\bibitem[Norsen(2002)]{Norsen2002}
Norsen T., 2002, Phys. Rev. C, 65, 045805
\bibitem[Norsen \& Reddy(2001)]{NorsenReddy2001}
Norsen T., Reddy S., 2001, Phys. Rev. C, 63, 065804
\bibitem[Oppenheimer \& Volkoff(1939)]{TOV}
Oppenheimer J. R., Volkoff G. M., 1939, Phys. Rev., 55, 374
\bibitem[Pandharipande \& Ravenhall(1989)]{FPS}
Pandharipande V.R., Ravenhall D.G., 1989,
Hot Nuclear Matter, in Nuclear Matter and Heavy Ion
Collision, NATO ADS Ser., vol. B205, ed. M. Soyeur
\bibitem[Pethick \& Ravenhall(1995)]{PethRav1995}
Pethick C.J., Ravenhall D.G., 1995, Annu. Rev. Nucl. Part.
Sci., 45,429
\bibitem[Pons et al.(2000)]{Pons2000}
Pons J.A., Reddy S., Prakash M., Lattimer J.M., 2000,
Phys. Rev. C, 62, 035803
\bibitem[Schaeffer, Haensel \& Zdunik(1983)]{SHZ1983}
Schaeffer R., Haensel P., Zdunik J.L., 1983, A\&A, 126, 121
\bibitem[Spyrou \& Stergioulas(2002)]{SpyrSterg2002}
Spyrou N.K., Stergioulas N., 2002, A\&A, 395, 151
\bibitem[Tolman(1939)]{Tolman39}
Tolman R.C., Phys. Rev., 55, 364
\bibitem[Weber(1999)]{weber.book}
Weber F., 1999, Pulsars as Astrophysical Laboratories for Nuclear
and Particle Physics, IoP Publishing, Bristol \& Philadelphia
\bibitem[Zdunik, Haensel \& Schaeffer(1987)]{fop2}
Zdunik J. L., Haensel P., Schaeffer R., 1987, A\&A, 172, 95
\end{thebibliography}
\end{document}